\documentclass[12pt,aps,nofootinbib,preprintnumbers,groupedaddress,letterpaper]{article}

\usepackage[dvips]{color}
\usepackage[normalem]{ulem}
\usepackage{amsmath}
\usepackage{enumerate}
\usepackage{amsfonts}
\usepackage{epsfig}
\usepackage{yfonts}
\usepackage{undertilde}

\usepackage{graphicx}
\usepackage{bm}
\usepackage{subfigure}

\newcommand\comment[1]{}

\newcommand\poincare{Poincar\' e }

\newcommand\ov{\over }

\def\le{\left}
\def\ri{\right}

\def\({\left(}
\def\){\right)}
\def\[{\left[}
\def\]{\right]}
\def\<{\langle}
\def\>{\rangle}
\newcommand\half{{\ensuremath{\frac{1}{2}}}}
\newcommand\p{\ensuremath{\partial}}

\newcommand\field[1]{{\ensuremath{\mathbb{{#1}}}}}

\newcommand{\RR}{\field{R}}

\newcommand{\be}{\begin{equation}}
\newcommand{\ee}{\end{equation}}
\newcommand{\bea}{\begin{eqnarray}}
\newcommand{\eea}{\end{eqnarray}}
\newcommand{\bwt}{\begin{widetext}}
\newcommand{\ewt}{\end{widetext}}

\newcommand{\bi}{\begin{itemize}}
\newcommand{\ei}{\end{itemize}}
\newcommand{\ben}{\begin{enumerate}}
\newcommand{\een}{\end{enumerate}}
\newcommand{\bca}{\begin{cases}}
\newcommand{\eca}{\end{cases}}
\newcommand{\bln}{\begin{align}}
\newcommand{\eln}{\end{align}}
\newcommand{\bst}{\begin{split}}
\newcommand{\est}{\end{split}}

\begin{document}

\begin{titlepage}

\begin{flushright}
QMUL-PH-21-02\\
\end{flushright}
\vskip 1cm

  \begin{center}
 
\centerline{\Large \bf {Relativistic membrane solutions in AdS$_4$}} 

\bigskip
\bigskip

{\bf David Vegh}

\bigskip

\small{
{ \it  Centre for Research in String Theory, School of Physics and Astronomy \\
Queen Mary University of London, 327 Mile End Road, London E1 4NS, UK}}

\medskip

{\it email:} \texttt{d.vegh@qmul.ac.uk}

\medskip

{\it \today}

\bigskip

%\date{\today}

\begin{abstract}

In this note we discuss various classical membrane solutions in AdS$_4$ spacetime: simple embeddings given by polynomials in ambient space, solutions with non-linear waves, and piecewise linear solutions.

\end{abstract}

\end{center}

\end{titlepage}

%\maketitle

\tableofcontents
\clearpage

%\comment{

%\clearpage

\section{Introduction}

The dynamics of classical Nambu-Goto strings in (2+1)-dimensional anti-de Sitter (AdS) spacetime can be described in many different ways. The system is integrable \cite{Pohlmeyer:1975nb, DeVega:1992xc, Bena:2003wd} which allows for an exact discretization of the equation of motion  \cite{Vegh:2015ska, Callebaut:2015fsa, Gubser:2016wno, Gubser:2016zyw, Vegh:2016hwq, Vegh:2016fcm}. Time-evolution can be computed by solving discrete equations which is advantageous because numerical errors do not arise. The corresponding string embedding is a piecewise linear {\it segmented string}.

A natural question is whether such constructions can be generalized to higher dimensional objects, for instance, to membranes moving in AdS$_4$. Even though in this case integrability techniques are not readily available, the membrane does `inherit' certain solutions which are higher-dimensional analogs of segmented strings. The purpose of this paper is to study these (and some other) solutions.

In the remainder of the Introduction, we describe the AdS geometry and the membrane action along with its equation of motion. In Section 2, we give a few simple membrane embeddings. The characterizing feature of these is that they are given by polynomial equations in ambient space (linear, quadric, cubic). Section 3 studies membrane solutions with non-linear waves traveling in one dimension. In Section 4 we describe how to attach linear patches. The attaching curve is a shockwave on the membrane where the membrane normal vector jumps. Section 5 and 6 discuss valence three and four vertices that may arise when different shockwaves cross.  Section 7 discusses the collision of three shockwaves on the brane.

%\clearpage

\subsection{Coordinate systems}

The canonical immersion of AdS$_4$ into the {ambient} $\RR^{2,3}$ linear space is obtained by considering the hyperboloid given by the equation
\be
  \label{eq:surface}
   X \cdot  X \equiv -X_{-1}^2 - X_0^2 + X_1^2 + X_2^2+ X_3^2 = -R^2  \qquad\qquad  X \in \RR^{2,3} .
\ee
Henceforth, we consider AdS spacetimes of unit radius and thus we set $R = 1$.
Global AdS wraps this surface infinitely many times and can be recovered by considering the universal cover of the hyperboloid: the angle on the $X_{-1}$, $X_0$ plane has to be `unwrapped' and this non-compact direction is then identified with global AdS time. The metric on AdS$_4$ is the induced metric in $\RR^{2,3}$ and its isometry group is $SO(2,3)$. In the following, vectors in the ambient (or embedding) space will be denoted by capital letters.

\clearpage

In the following, we will also use  coordinates which only cover the \poincare patch. They will be called {\it \poincare coordinates} and can be defined by
\be
  \nonumber
  (t, \, x, \, y, \, z) = \le(  {X_{0} \over X_{3} - X_{-1}}, \ {X_{1} \over X_{3} - X_{-1}}, \ {X_{2} \over X_{3} - X_{-1}}, \ {1 \over X_{3} - X_{-1}} \ri).
\ee
The metric written in \poincare coordinates is clearly conformally flat
\be
  \label{eq:poinca}
  ds^2 = {-dt^2 + dx^2  + dy^2 + dz^2 \over z^2}
\ee
Gravity `acts' in the $z$ direction. Massive objects never reach the boundary at $z=0$. They fall `down' in the positive $z$ direction.

Another coordinate system $\{\tau, r, \theta, \phi \}$ is obtained by the transformation
\be
   X =  \left( \begin{array}{c}
     \cosh r \cos \tau   \\
     \cosh r \sin \tau   \\
     \sinh r \sin \theta \sin \phi   \\
     \sinh r \sin \theta \cos \phi   \\
     \sinh r \cos \theta
   \end{array} \right)
\ee
As mentioned before, $\tau$ is the (non-compact) global time direction.
It is easy to see that this parametrization satisfies (\ref{eq:surface}).
In these coordinates, the metric becomes
\be
  \label{eq:globalmetric2}
  ds^2 = -\cosh^2 r \, {d\tau}^2+{dr}^2+\sinh^2 r \left({d\theta}^2+{d\phi}^2 \sin^2 \theta \right)
\ee

For purposes of visualization, a somewhat nicer radial coordinate is  $\rho \in [0,1)$ defined by
\be
  \nonumber
  \rho = \sqrt{\cosh r - 1 \over \cosh r +1}
\ee
The metric in terms of this coordinate is
\be
  \nonumber
  ds^2 = -\le({1+\rho^2 \over 1-\rho^2} \ri)^2  \, d\tau^2 + {4  \over (1-\rho^2)^2} \le(  d\rho^2 + \rho^2 \, d\Omega^2 \ri)
\ee
where $d\Omega^2 \equiv {d\theta}^2+{d\phi}^2 \sin^2 \theta$. A timeslice of global AdS  is then just a ball of unit radius. We will call these variables {\it global coordinates}.

\clearpage

\subsection{Action and equation of motion}

Our starting point will be the (2+1)-dimensional membrane action %\footnote{For other brane actions see e.g. \cite{Schmidhuber:1996fy}.}
coupled to a background three-form field $C$
\be
  \label{eq:action}
  S  = -\int  d^3 \sigma \sqrt{-g} +  \kappa \int d^3 \sigma \, \p_a Y^M \p_b Y^N\p_c Y^L \epsilon^{abc} C_{MNL}
\ee
Since we will be considering classical solutions only, for simplicity, we have rescaled the action such that the brane tension is set to one.
Here $\sigma^\mu$ and $Y^{M}$ are coordinates on the worldvolume and AdS$_4$, respectively. The induced metric on the brane is denoted $g_{\mu\nu}$. We choose a background three-form field such that its field strength $dC$ is the volume form of AdS$_4$. This choice preserves all the symmetries of the target space. $\kappa$ parametrizes the strength of the coupling of the membrane to the pull-back of the three-form.

If we use \poincare coordinates for AdS, we have $\{ Y^M \} = \{t, x, y, z\}$. Worldvolume coordinates can be chosen such that $\{\sigma^1, \sigma^2, \sigma^3 \} = \{t, x, y \} $. The embedding will be characterized by a single function $z(t,x,y)$ which gives the brane position in the $z$ (radial) direction. On the \poincare patch, the antisymmetric three-form can be chosen
\be
  C = {dt \wedge dx \wedge dy \over  z^{3}} .
\ee
The equation of motion is
\be
  \label{eq:eom}
  (1+ z_{,\alpha}z^{,\alpha}) {z_{,\beta}}^{,\beta} - z^{,\alpha}z^{,\beta}z_{,\alpha\beta}
  + {3\ov z} (1+ z_{,\alpha}z^{,\alpha}) \pm {3 \kappa\ov z} (1+ z_{,\alpha}z^{,\alpha})^{3\ov 2}= 0
\ee
\comment{
\bea
  \nonumber
  0 & = &  -\le[ 1 + (\p_x z)^2 + (\p_y z)^2   \ri] \p_t^2 z
  + \le[ 1 - (\p_t z)^2 + (\p_y z)^2   \ri] \p_x^2 z
  + \le[ 1 - (\p_t z)^2 + (\p_x z)^2   \ri] \p_y^2 z \\
  \nonumber
  &&
  + 2\le[  -(\p_x\p_y z) (\p_x z) (\p_y z)  + (\p_t\p_y z) (\p_t z) (\p_y z) + (\p_t\p_x z) (\p_t z) (\p_x z) \ri] \\
  && +{3 \ov z} \le[ 1 - (\p_t z)^2+(\p_x z)^2+(\p_y z)^2    \ri] + {3 \kappa \ov z} \le[  1 - (\p_t z)^2+ (\p_x z)^2 + (\p_y z)^2\ri]^{3/2}
\eea
}
where $\alpha, \beta \in \{ t,x,y \}$, comma indicates partial derivative (e.g. $z_{,\alpha\beta} \equiv \p_\alpha \p_\beta z$), and indices are lowered/raised using a flat metric $\eta = \textrm{diag}(-1,1,1)$. The first two terms would be the equation of motion in Minkowski target space. In AdS$_4$, these are modified by the presence of the third term. The sign of the last term depends on whether the surface is in a normal position or `upside-down'. (This can happen when the solution is a double-valued function $z=z_\pm(t,x,y)$. In this case, the sign will be different for the two branches.)

In the following, we will be interested in solutions to this non-linear partial differential equation.

\clearpage

\begin{figure}[h]
\begin{center}
\includegraphics[width=6cm]{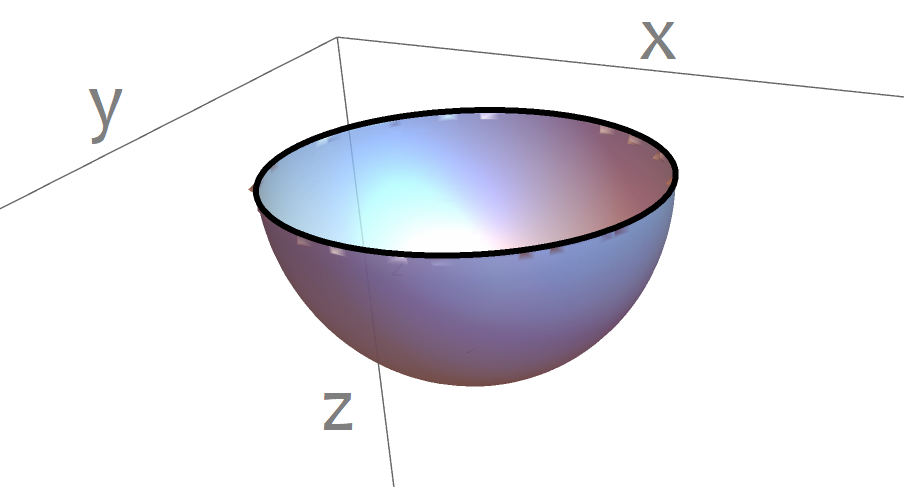}
\caption{\label{fig:onepatch} Linear solution with $\kappa = 0$. The shape of the brane at any given time on the \poincare patch is a hemisphere. The equator lies on the boundary of AdS. The radius $R$ of the hemisphere changes in time according to $R^2 = R_0^2 + (t-t_0)^2$.
}
\end{center}
\end{figure}

\section{Simple membrane solutions in AdS$_4$ }

In this section we are going to discuss a few simple solutions to the membrane equation of motion in (\ref{eq:eom}). These solutions can be defined in the $\RR^{2,3}$ ambient space by a polynomial equation. A characterizing feature is the degree of the (homogeneous) polynomial. In the following subsections, we list classes of such solutions according to their degree.

%\subsection{AdS$_3$ patches}
\subsection{Linear surfaces}
\label{sec:lin}

The simplest solution lies in a linear subspace in the ambient space,
\be
  \label{eq:linsol}
  X \cdot N = \lambda
\ee
where $X, N \in \RR^{2,3}$. Here $X$ is a point in the embedding, $N$ is a fixed unit-length vector and $\lambda$ is a constant. This surface has to be intersected with the hyperboloid in (\ref{eq:surface}). The resulting three-dimensional subspace is the worldvolume of the membrane.

It is easy to see that on the \poincare patch (\ref{eq:linsol}) translates into either (i) planes at a fixed angle
\be
  z(t,x,y) = {x\ov \lambda} ,
\ee
or (ii) shrinking and expanding spheres
\be
  z(t,x,y) = \lambda + \sqrt{1+ t^2-x^2-y^2+\lambda^2} .
\ee
and their shifted, rotated, and boosted cousins.

The equation of motion in (\ref{eq:eom}) is satisfied if
\be
  \lambda =   \pm{\kappa \ov \sqrt{1-\kappa^2}}
\ee
For $\kappa = 0$ this gives $\lambda = 0$ and $N$ is the constant {\it normal vector} of the membrane. For $\kappa \ne 0$, $N$  may be called a {\it generalized normal vector}.

The spheres are centered on the boundary if $\kappa = 0$. An example is shown in Figure \ref{fig:onepatch}. The three-dimensional picture is a timeslice of the \poincare patch. At any given instance the shape of the membrane is a hemisphere. The black circle (the equator) lies on the $z=0$ boundary of AdS.

\clearpage

\subsection{Quadric surfaces}

Certain quadric surfaces (a.k.a. `hypercycles') in the ambient space also satisfy the equation of motion. In the following, we give examples for such curves in $SO(2,3)$ frames in which the defining equations are the simplest.

\begin{figure}[h]
\begin{center}
\includegraphics[width=4cm]{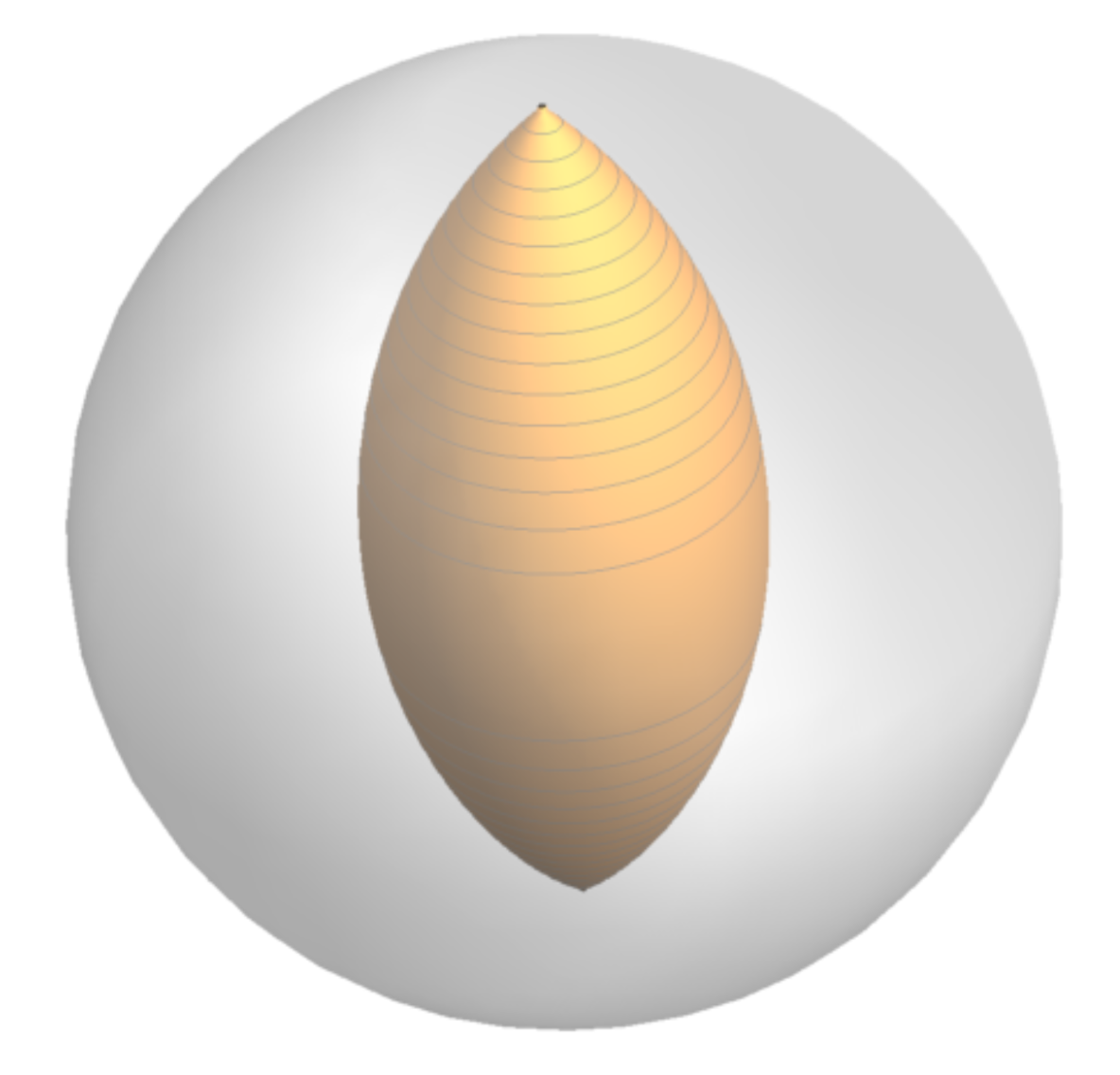}
\caption{\label{fig:cone} A quadric membrane solution on a timeslice of global AdS$_4$. The (yellow) membrane connects two antipodal points on the boundary of AdS (i.e. the surface of the ball). For such a static solution to exist, the coupling  to the three-from field has to be non-zero. This provides an extra force on the membrane and keeps it from collapsing.
}
\end{center}
\end{figure}

\begin{itemize}

\item
There are degree-2 curves which in a frame have the form
\be
  X_1^2 + X_2^2 = \lambda
\ee
where $\lambda>0$.
On the \poincare patch we get
\be
   z(t,x,y) =\sqrt{x^2 + y^2 \ov \lambda}
\ee
This is a nothing but a static cone. It can be thought of as a `fat string' which hangs from the boundary point at $x=y=z=0$. In global coordinates, the brane has a cigar shape as seen in Figure \ref{fig:cone}.

The embedding satisfies the equation of motion if $\lambda$ can be set such that
\be
  \kappa = -{\lambda +{1\ov 3} \ov \sqrt{\lambda(1+\lambda)}}
\ee
Note that there is no surface corresponding to $\kappa=0$.

\item Another curve is given by
\be
  X_{-1}^2 + X_2^2 = \lambda
\ee

For the embedding on the \poincare patch we obtain two solutions
\be
  z_\pm(t,x,y) = \sqrt{t^2-x^2-y^2-1 -2 \lambda \pm 2 \sqrt{ (1+\lambda)x^2 + \lambda(1-t^2+y^2+\lambda)}}
\ee

which satisfy the equation of motion if $\lambda$ is set such that the equation
\be
  \kappa =  \mp{\lambda +{1\ov 3} \ov \sqrt{\lambda(1+\lambda)}}
\ee
is satisfied.
Assuming a fixed positive sign in (\ref{eq:eom}), here minus sign should be taken for the $z_+$ solution and plus sign for $z_-$.

\item Yet another curve is given by
\be
  X_{-1}^2 + X_{0}^2 = \lambda
\ee
The \poincare embedding is
\be
  z_\pm(t,x,y) = \sqrt{t^2-x^2-y^2-1 -2 \lambda \pm 2 \sqrt{ -(1+\lambda)t^2 + \lambda(1+x^2+y^2+\lambda)}}
\ee

\end{itemize}

\begin{figure}[h]
\begin{center}
\includegraphics[width=5.5cm]{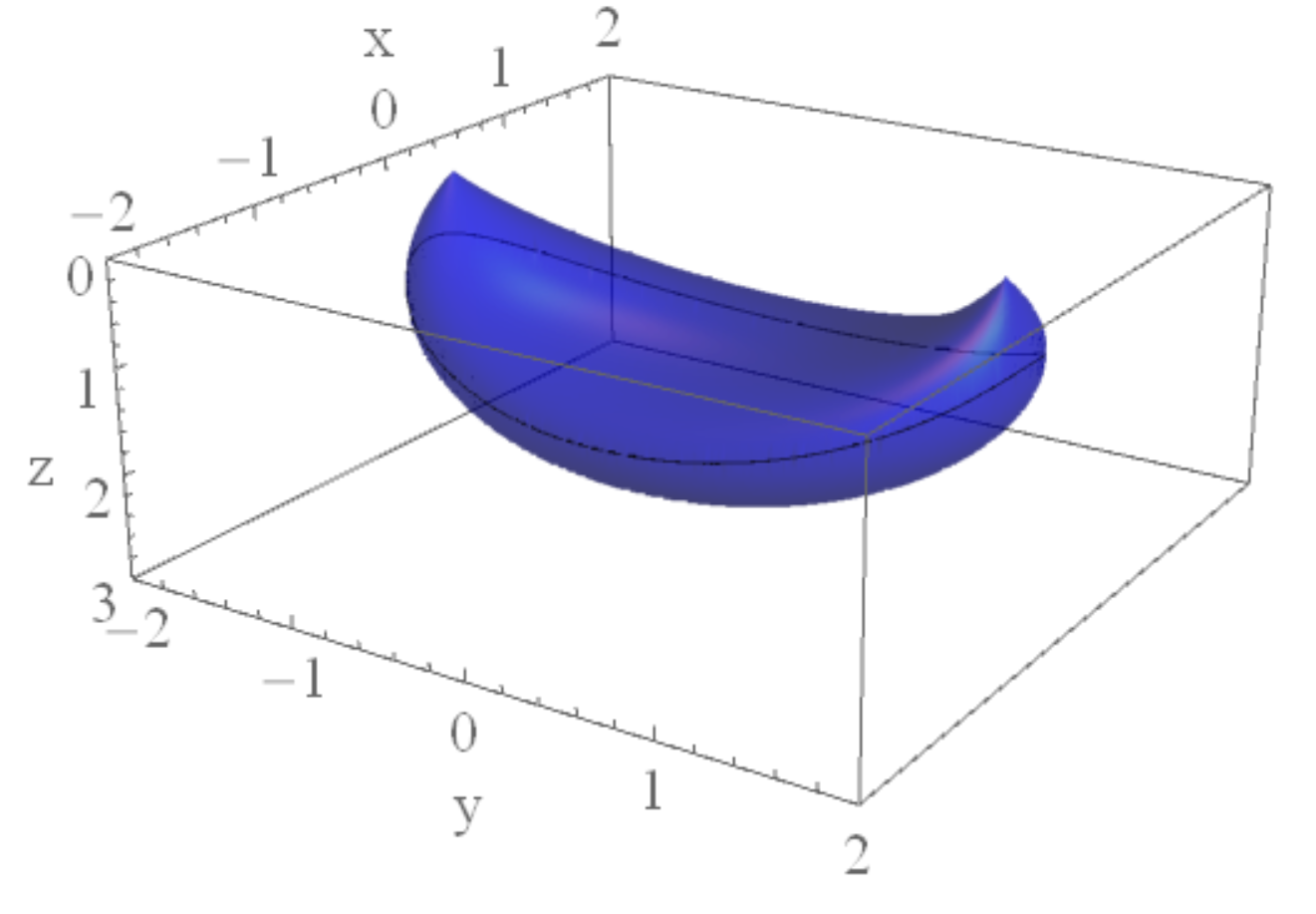}
\includegraphics[width=5.5cm]{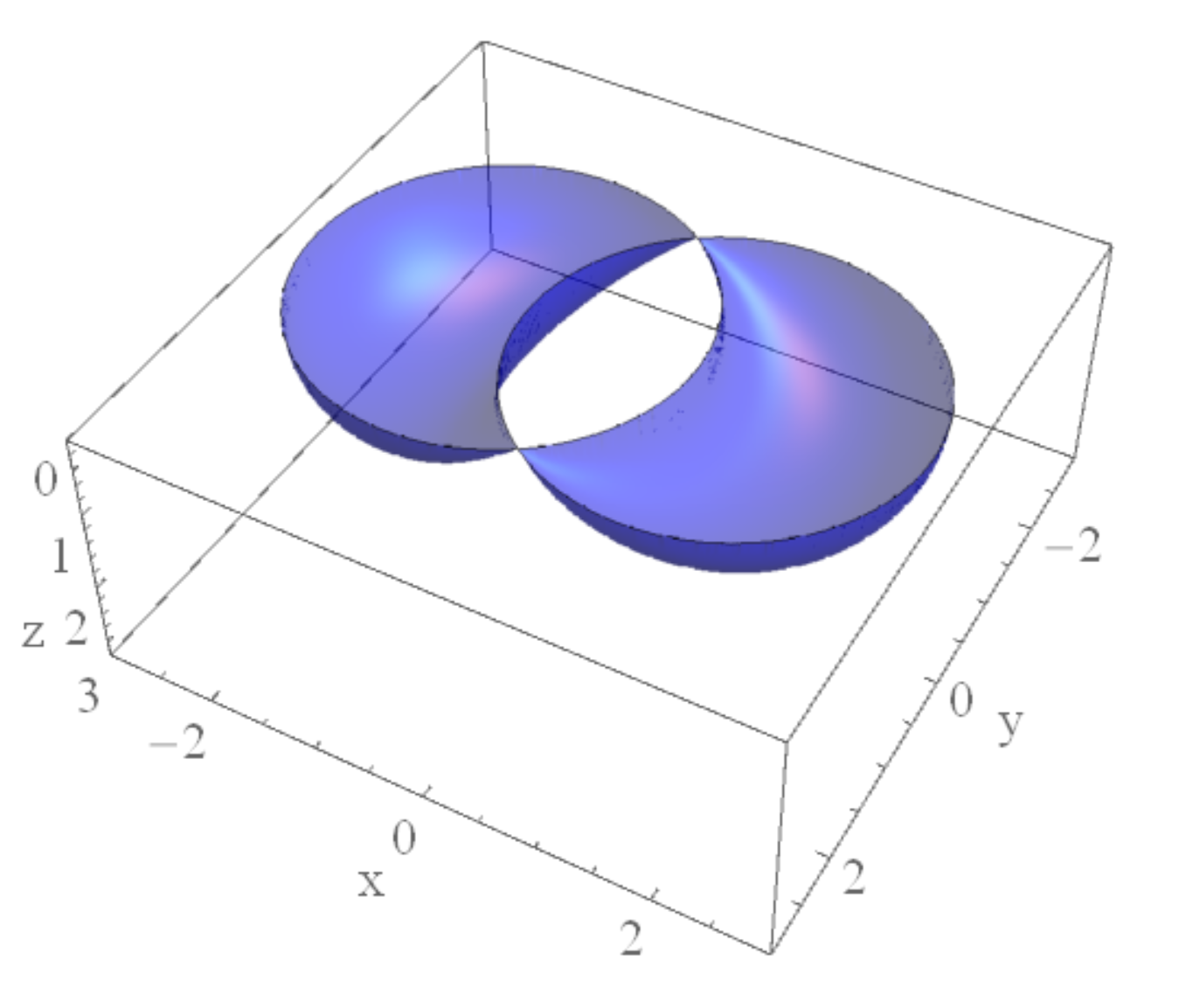}
\includegraphics[width=5.0cm]{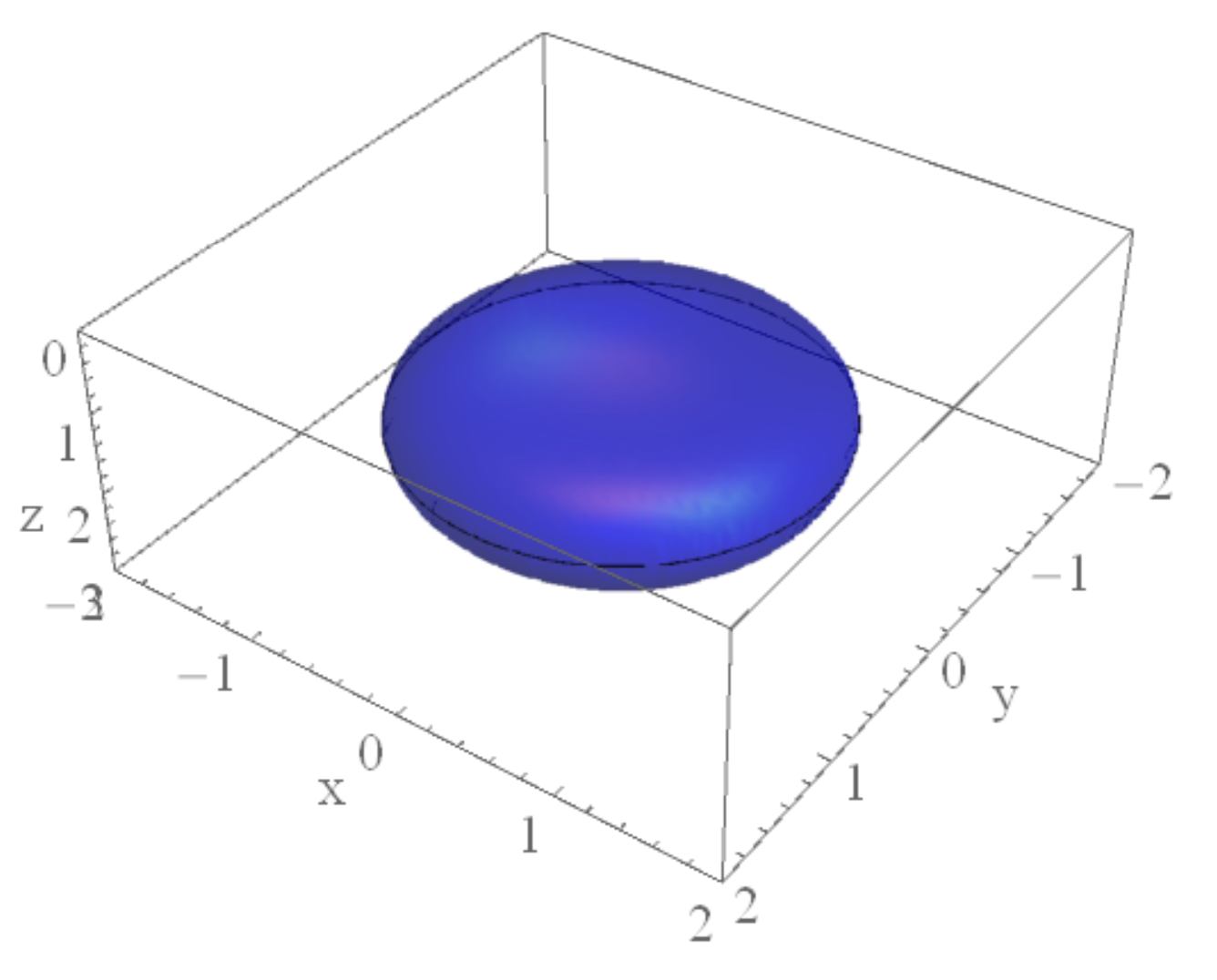}
\caption{\label{fig:quadric3}
Quadric membrane solutions on the \poincare patch. The top of the bounding boxes are the $z=0$ boundary planes of AdS.
}
\end{center}
\end{figure}

Examples for these solutions are plotted in Figure \ref{fig:quadric3}.

\clearpage

\begin{figure}[h]
\begin{center}
\includegraphics[width=4cm]{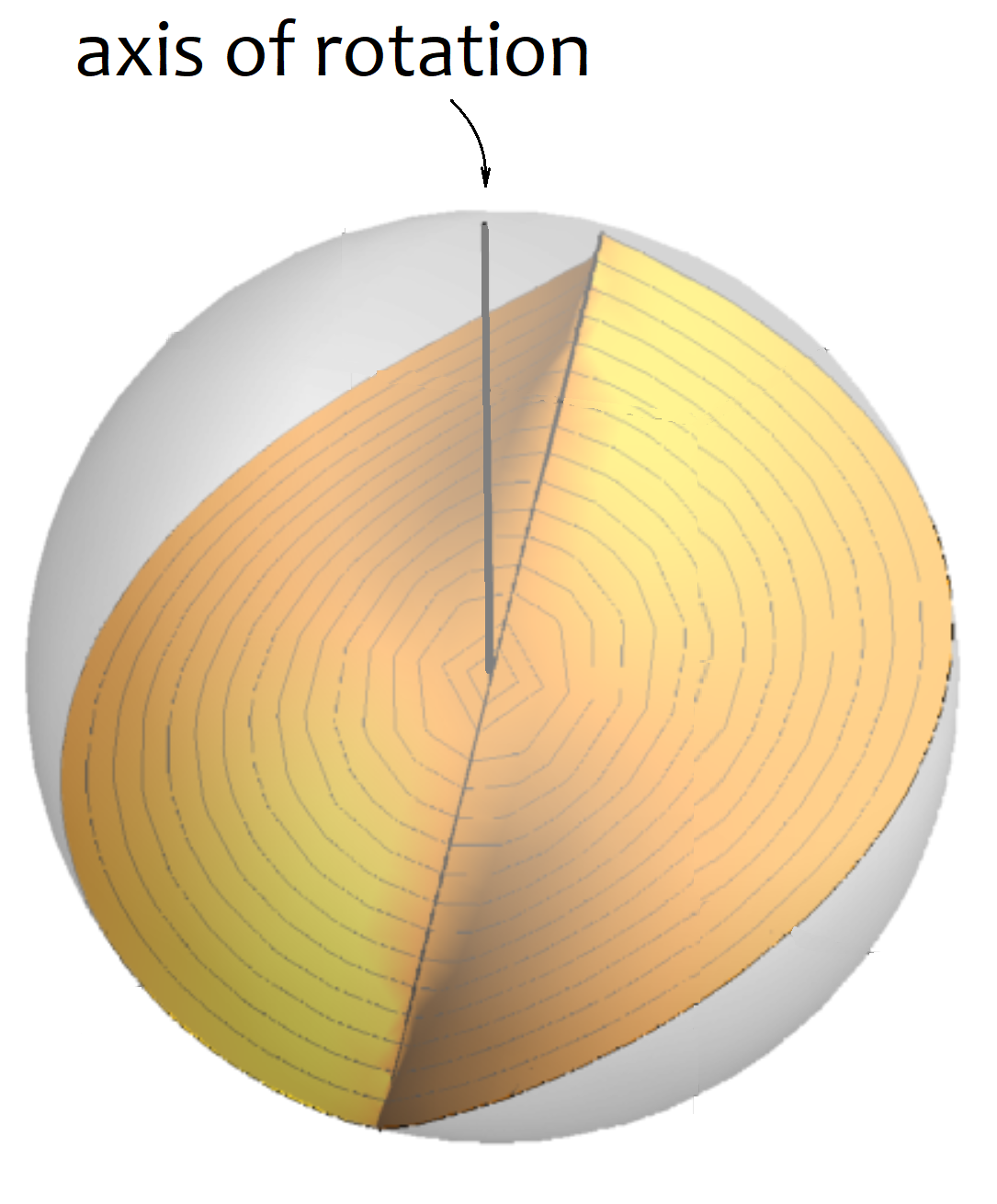}
\caption{\label{fig:cubicglobal} A timeslice of the cubic solution in global AdS. The (yellow) membrane rotates about the vertical axis as a rigid body.
}
\end{center}
\end{figure}

%\vskip -1.5cm \quad
\subsection{Cubic surfaces}

\begin{figure}[h]
\begin{center}
\includegraphics[width=5cm]{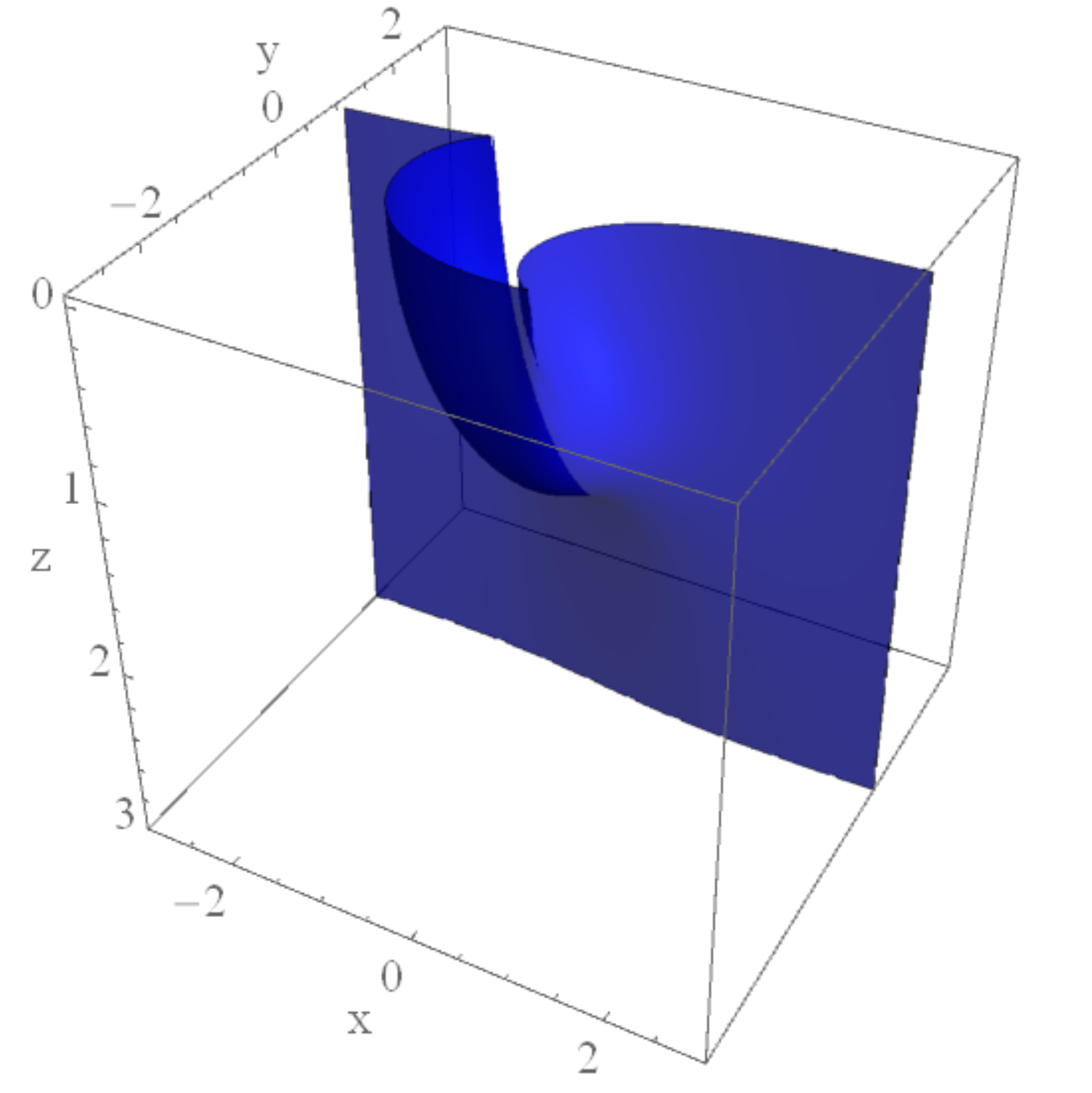}
\includegraphics[width=5cm]{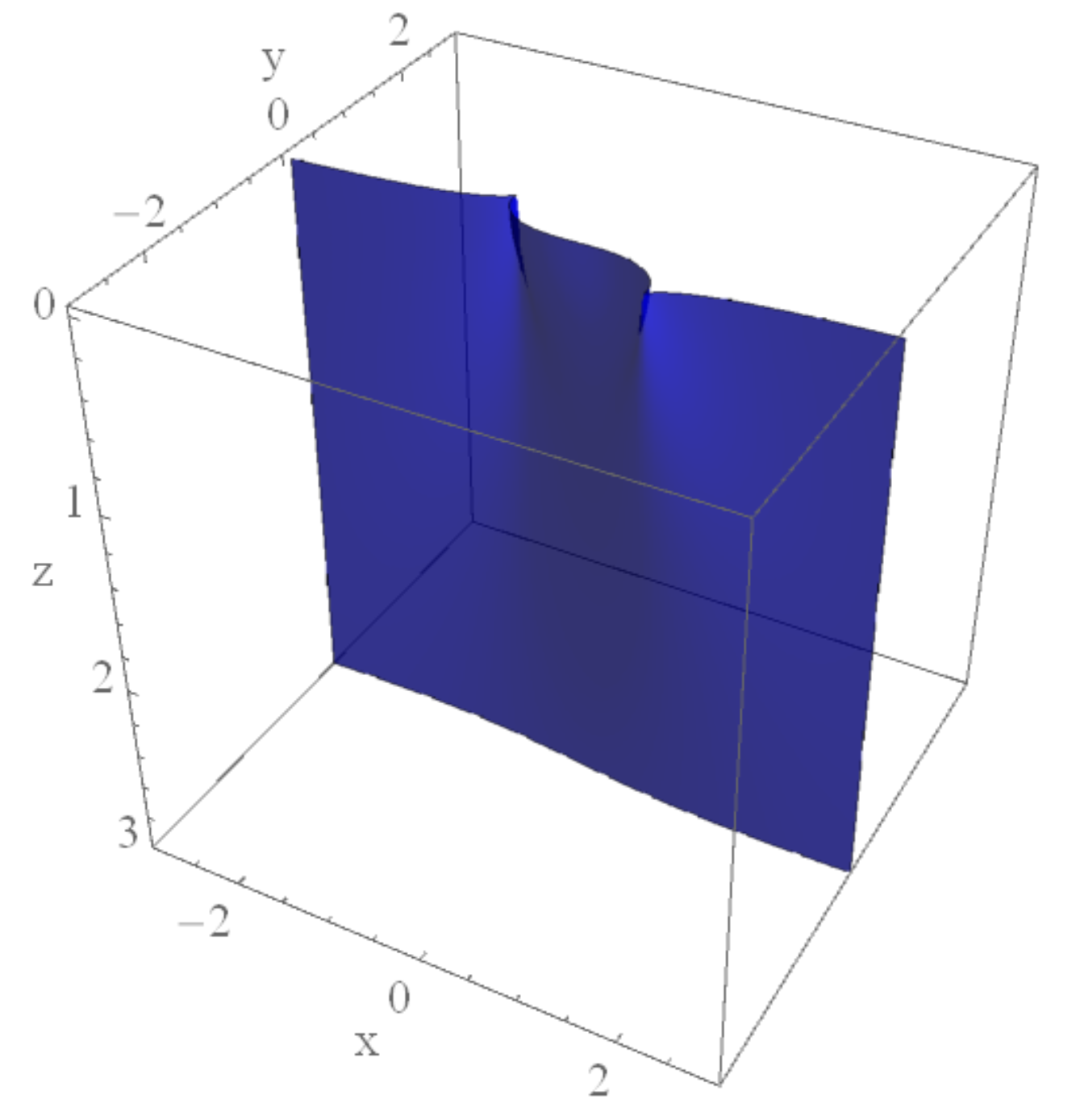}
\includegraphics[width=5cm]{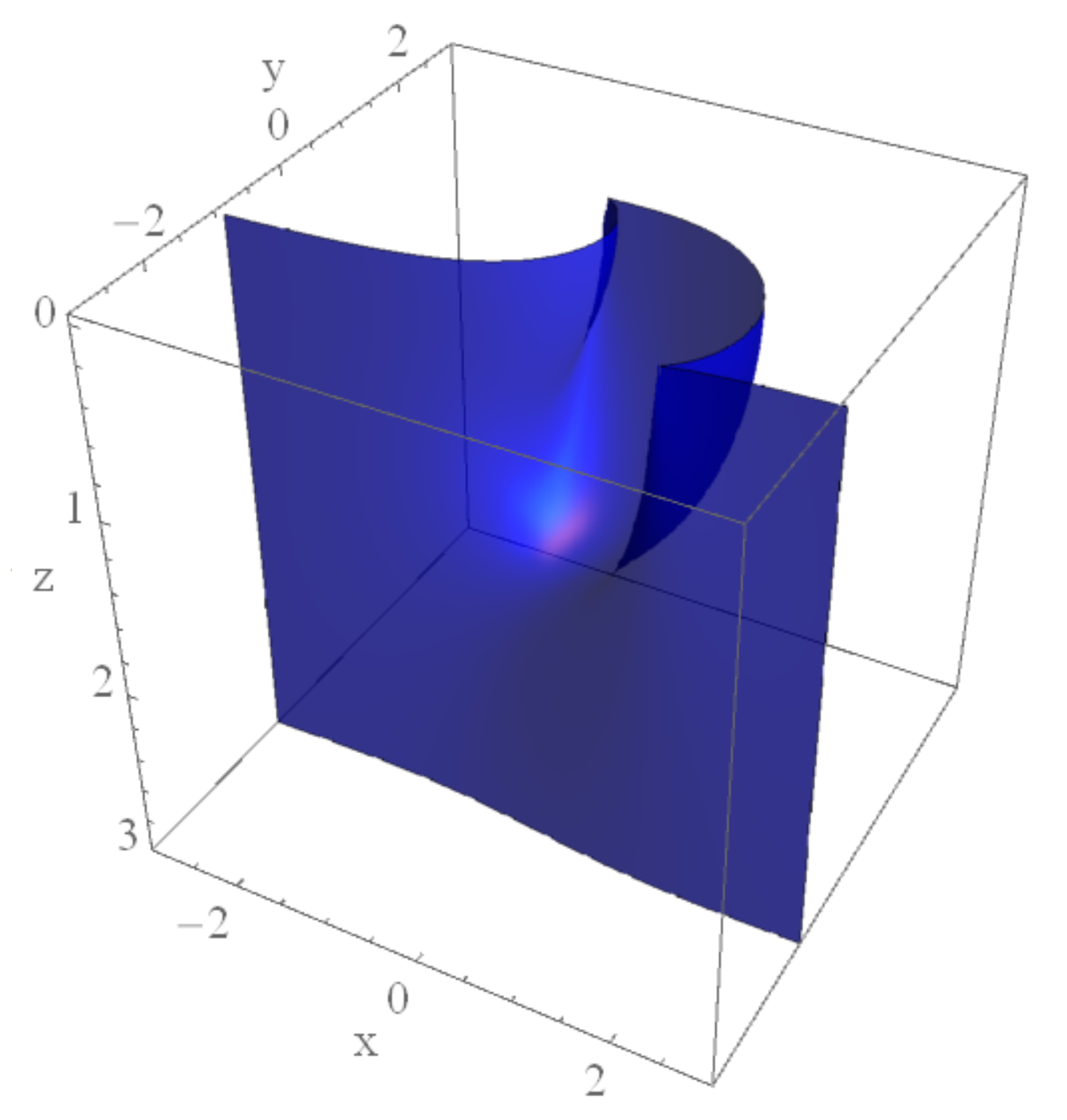}
\caption{\label{fig:cubic}
Time-evolution of the cubic membrane solution on the \poincare patch. The top of the bounding box is the boundary of AdS. Here the membrane has two cusps.
}
\end{center}
\end{figure}

A cubic solution to the EOM with $\kappa=0$ is given by the (Wick-rotated) Cartan-M\" unzner polynomial
\be
  X_{3}^3 -3\sqrt{3} X_{-1}X_0 X_2 + {3\sqrt{3}\ov 2} X_1(X_{-1}^2 - X_0^2) - {3\ov 2} X_3 \le[ X_{-1}^2 + X_0^2 + 2 (X_1^2 + X_2^2) \ri] = 0
\ee
In global AdS coordinates the embedding is given by\footnote{The equation is linear in $R \equiv \tanh^2 r$. Using this variable as a new radial coordinate, we can parametrize membranes via $R=R(\tau, \theta, \phi)$. The equation of motion with no coupling to the three-form turns out to be
\bea
   \nonumber
  &&
   0 =  4 \sin ^3(\theta ) \left(\p_t^2 R +2\right) R^4-2 \sin (\theta ) \p_\phi R  \p_\theta R
   \p_\theta\p_\phi R -\sin (\theta ) R \left(\left(\p_t^2 R +8\right) \p_\phi R ^2-2
   \p_t R  \p_t\p_\phi R  \p_\phi R  \ri. \\
   \nonumber
  &&
   \le. + \p_t R ^2 \left( \p_\phi^2 R +\sin ^2(\theta ) \p_\theta^2 R \right)+\sin ^2(\theta ) \p_\theta R ^2 \left(\p_t^2 R +8\right)+\sin
   (\theta ) \p_\theta R  \p_t R  \left(\cos (\theta ) \p_t R -2 \sin (\theta ) \p_t \p_\theta R\right)\right)    \\
   \nonumber
   &&
     -2 \sin (\theta ) R^3 \left(2 \p_\phi^2 R +\sin (\theta ) \left(\sin (\theta )
   \left(\p_t R ^2+2 \p_\theta^2 R +2 \p_t^2 R +8\right)+2 \cos (\theta ) \p_\theta R \right)\right) \\
   \nonumber
   &&
   +2 \sin (\theta ) R^2 \left(2 \p_\phi R ^2+2 \p_\phi^2 R +\sin (\theta ) \left(\sin
   (\theta ) \left(3 \p_t R ^2+2 \left(\p_\theta R ^2+\p_\theta^2 R \right)\right)+2 \cos (\theta )
   \p_\theta R \right)\right) \\
   \nonumber
   &&
   +\sin (\theta ) \p_\theta R ^2 \left(\p_\phi^2 R +\sin (\theta ) \cos (\theta )
   \p_\theta R \right)+\p_\phi R ^2 \left(\sin (\theta ) \p_\theta^2 R +2 \cos (\theta ) \p_\theta R \right)+8 \sin ^3(\theta ) R^5
\eea
It can be checked that the cubic solution does indeed satisfy this equation.
}
\be
  2\cos 3\theta \tanh^2 r - 3\cos \theta - 3\sqrt{3}\sin \theta \sin(2\tau-\phi) = 0
\ee
Since $\tau$ and the angle $\phi$ only appear through the combination $2 \tau-\phi$, the surface rotates with a constant angular velocity like a rigid body. This is depicted in Figure \ref{fig:cubicglobal}. Figure \ref{fig:cubic} shows the time-evolution of the solution on the \poincare patch.
The membrane touches the boundary along a curve which contains two cusps.

\clearpage

\begin{figure}[h]
\begin{center}
\includegraphics[width=7cm]{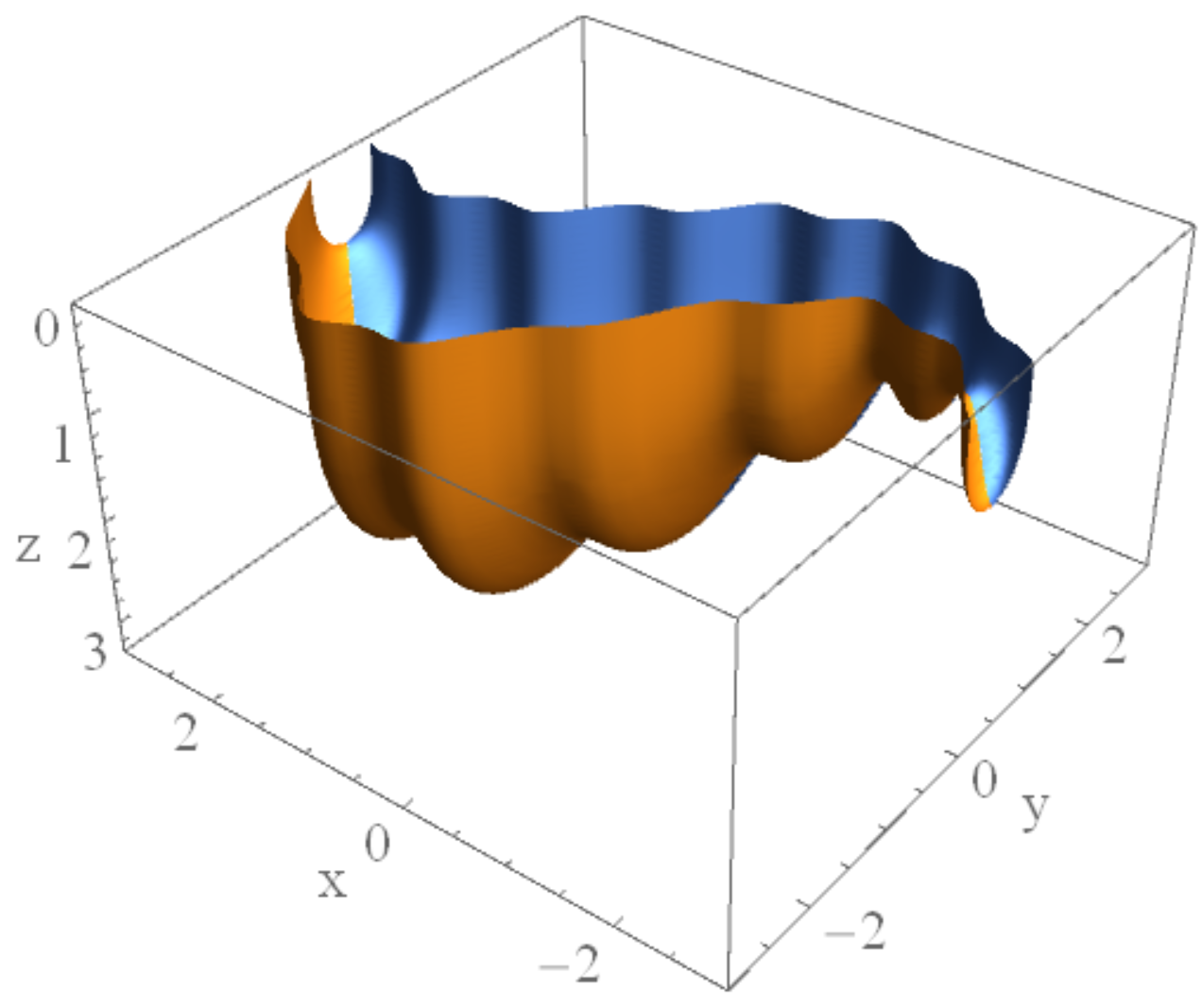}
\caption{\label{fig:waves} Traveling waves on a brane embedded into the \poincare patch. The picture shows a timeslice. The top of the bounding box is the AdS boundary.
}
\end{center}
\end{figure}

\section{Waves traveling in one direction }

Instead of parametrizing the brane embedding on the \poincare patch via $z(t,x,y)$, one can find the equation of motion for $y(t,x,z)$. Let us take the wave ansatz
\be
  y(t,x,z) = p(x^-, z)  \qquad \textrm{where} \quad x^- \equiv t-x
\ee
The equation of motion simplifies
\be
  \label{eq:waveeqn}
  \p_z p + (\p_z p)^3 - {z\ov 3} \p_z^2 p + \kappa \le[ 1+ (\p_z p)^2  \ri]^{3\ov 2} = 0
\ee
Note that the equation only contains $z$-derivatives. Let us first set $\kappa=0$. Then the solution is given by
\be
   p(x^-, z) = {z^4 \ov 4 A(x^-)^3 } \, {}_{2}F_{1}\le({1\ov 2}, \, {2\ov 3}; \, {5\ov 3}; \, {z^6 \ov  A(x^-)^6}\ri) + B(x^-)
\ee
with $A$ and $B$ are two arbitrary functions.

Let us shift $B$
\be
  B_\pm(x^-) :=  \pm {\sqrt{\pi} \, \Gamma({5\ov 3}) \ov 4 \Gamma({7\ov 6}) } A(x^-) + C(x^-)
\ee
Then the two solutions
\be
   p_- = {z^4 \ov 4 A^3 } \, {}_{2}F_{1}\le( {1\ov 2}, \, {2\ov 3}; \, {5\ov 3}; \, {z^6 \ov  A^6} \ri) +  B_-
\ee
and
\be
   p_+ = -{z^4 \ov 4 A^3 } \, {}_{2}F_{1}\le({1\ov 2}, \, {2\ov 3}; \, {5\ov 3}; \, {z^6 \ov  A^6}\ri) +  B_+
\ee
connect smoothly. The membrane touches the boundary along two curves which are parametrized by $A$ and $C$. The curves move with the speed of light on the boundary and the membrane follows their motion as a rigid body. A snapshot of such a solution on the \poincare patch is shown in Figure \ref{fig:waves}.

At non-zero $\kappa$, the equation of motion (\ref{eq:waveeqn}) can be solved in terms of  Appell hypergeometric functions
\bea
  \nonumber
  && p(x^-, z) = -{z \kappa \ov \sqrt{1-\kappa^2} } \, F_1\le( {1\ov 3}; \, {1\ov 2}, \, {1\ov 2}; \, {4\ov 3}; \, {z^3 \ov (\kappa+1)A(x^-)^3}, \, {z^3 \ov (\kappa-1)A(x^-)^3}   \ri) \\
  \nonumber
  && + {z^4 \ov 4 A(x^-)^3 \sqrt{(1-\kappa^2)}}
  \, F_1\le( {4\ov 3}; \, {1\ov 2}, \, {1\ov 2}; \, {7\ov 3}; \, {z^3 \ov (\kappa+1)A(x^-)^3}, \, {z^3 \ov (\kappa-1)A(x^-)^3}   \ri)  + B(x^-)
\eea
where  $A$ and $B$ are again two arbitrary functions.

\clearpage

\begin{figure}[h]
\begin{center}
\includegraphics[width=10cm]{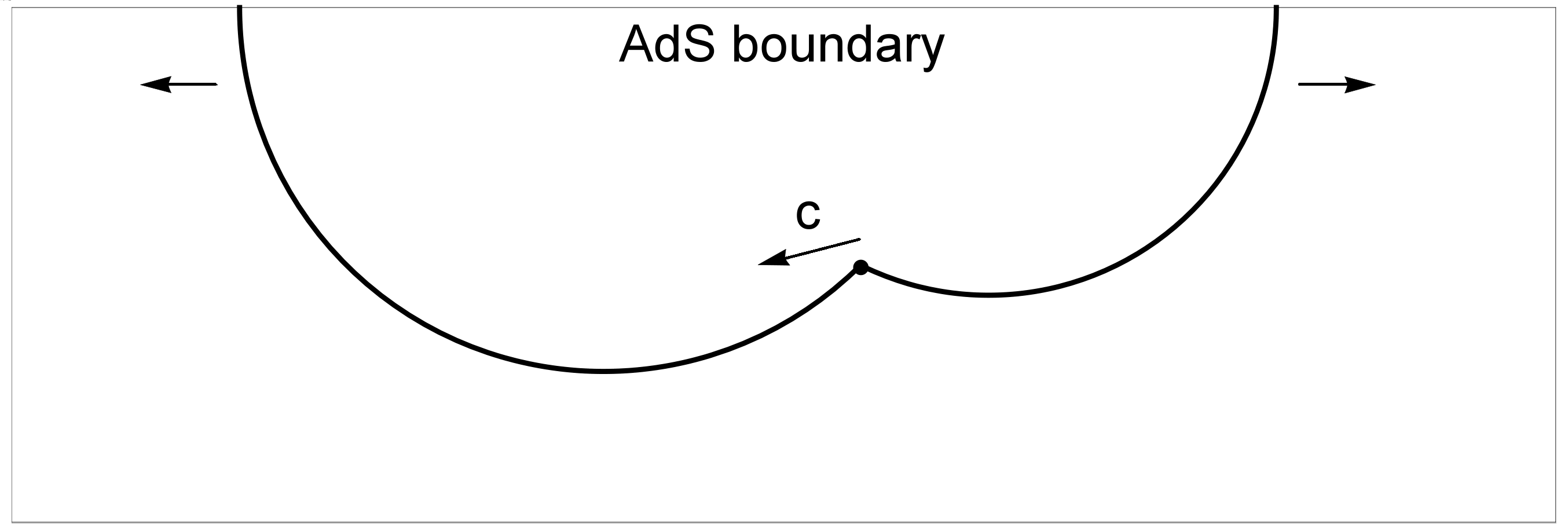}
\caption{\label{fig:string} Segmented string on the \poincare patch of AdS$_3$ \cite{Vegh:2015ska}. The string consists of two patches (arcs) which have constant normal vectors. The kink between the two segments moves with the speed of light.
}
\end{center}
\end{figure}

\section{Attaching patches}

Segmented strings in flat space are piecewise linear solutions. Kinks between the segments move with the speed of light and their worldlines form a rectangular lattice on the string worldsheet.  This idea can be generalized to AdS$_3$ where the embedding is built from AdS$_2$ patches (see \cite{Vegh:2015ska, Callebaut:2015fsa})\footnote{For recent developments, the reader is referred to \cite{Vegh:2016hwq, Gubser:2016wno, Gubser:2016zyw, Vegh:2018dda}.}. The segmented string construction provides an exact discretization of the non-linear string equations of motion.
 In Figure~\ref{fig:string} we show an example which consists of two elementary patches (arcs). Any smooth string can be approximated by segmented strings to arbitrary accuracy by increasing the number of segments.

In this section we would like to generalize these ideas for the case of two-branes in AdS$_4$. The membrane has one transverse direction and therefore in certain cases its classical dynamics is similar to that of a string in AdS$_3$.

For simplicity, we henceforth set $\kappa = 0$. Let us consider two linear brane patches (see section~\ref{sec:lin}) with constant normal vectors $N_1, N_2 \in \RR^{2,3}$. At a given time, the two patches may intersect along a curve. This curve is analogous to the kink in Figure~\ref{fig:string}. We want the worldsheet of this curve to be a null surface in  AdS$_4$ so that it is a shockwave on the membrane, propagating with the speed of light. In the case of a string, this happens precisely when the normal vectors (which are elements of $\RR^{2,2}$) satisfy the compatibility condition \cite{Vegh:2015ska}
\be
  \label{eq:compat}
  N_1\cdot  N_2  = 1
\ee
In this case the kink moves with the speed of light linearly in the usual \poincare coordinate system. We will see that the same condition applies for the case of two membrane patches.

\begin{figure}[h]
\begin{center}
\includegraphics[width=10cm]{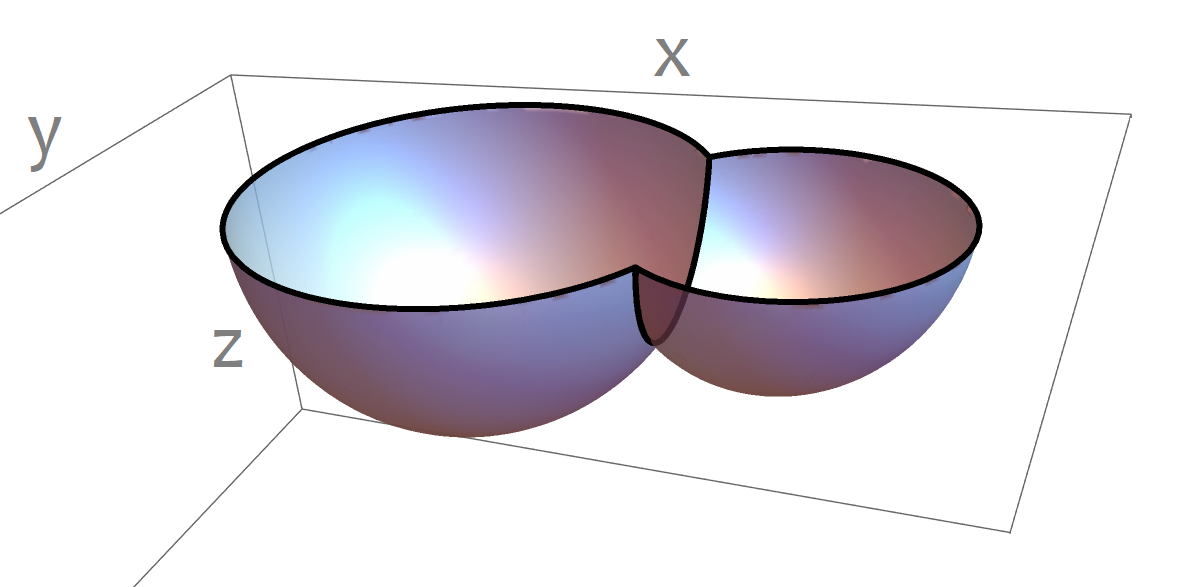}
\caption{\label{fig:twopatches}
Two compatible linear patches on the \poincare patch. The vertical black arc is a shockwave which travels with the speed of light. The construction is a higher-dimensional analog of the segmented string in Figure~\ref{fig:string}.
}
\end{center}
\end{figure}

Let us first apply an $SO(2,3)$ transformation such that the unit normal vectors are rotated into a frame in which
\bea
  \nonumber
  N_1 &=& (0, \, 0, \, 0, \, 0, \, 1)^T \\
  \nonumber
%  N_2 &=& (a, \,  0, \,  0, \,  a+\epsilon,  \, \sqrt{1+a^2-(a+\epsilon)^2} )^T
  N_2 &=& (a, \,  0, \,  0, \,  a+\epsilon,  \, \sqrt{1-2a\epsilon -\epsilon^2} )^T
\eea
The intersection locus  is two-dimensional in spacetime. Points on it satisfy
\be
  X\cdot N_{1,2} = 0 \qquad X^2 = -1
\ee
This is solved by
{\footnotesize
\be
 \nonumber
 X = (\sigma_1,\frac{\sqrt{a^2 \left(\sigma_2^2+1\right)+\epsilon(2 a +\epsilon ) \left(1-\sigma_1^2+\sigma_2^2\right) }}{a+\epsilon },\sigma_2,\frac{a \sigma_1}{a+\epsilon },0)
\ee
}
where $\sigma_{1,2}$ parametrizes the worldsheet of the shockwave.
The induced metric on it is given by
\be
  \nonumber
  \det g = -{\epsilon (2 a + \epsilon) \over a^2 (1+\sigma_2^2) + \epsilon (1-\sigma_1^2 + \sigma_2^2) (2a+\epsilon)}
\ee
Clearly, the shockwave defined by $X$ moves everywhere with the speed of light iff $\epsilon=0$. (Here we have assumed that $a \ne 0$. If $a=0$ then the two normal vectors are not compatible.)
Note that $\epsilon=0$ is equivalent to $N_1\cdot  N_2  = 1$.
This formula is invariant under $SO(2,3)$ and thus it is valid for branes in general positions.
We will call adjacent brane patches that satisfy this constraint {\it compatible}\footnote{The compatibility condition generalizes to higher-dimensional codimension one branes as well.}.

The membrane analog of Figure~\ref{fig:string} is depicted in Figure~\ref{fig:twopatches}.

\clearpage

\begin{figure}[h]
\begin{center}
\includegraphics[width=3cm]{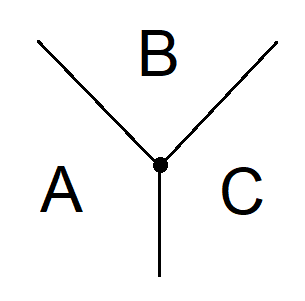}
\caption{\label{fig:tri1}
Three compatible linear patches with normal vectors $A$, $B$, and $C$. The lines represent the shockwaves.
}
\end{center}
\end{figure}

\section{Trivalent vertex on the boundary}

Once there are shockwaves on the brane, we need to understand what happens when they collide. For simplicity, let us consider linear patches again with $\kappa=0$. Is the trivalent vertex of shockwaves in Figure \ref{fig:tri1} possible? We assume that the patches are pairwise compatible, i.e.
\be
  A \cdot B = A \cdot C = B \cdot C = 1
\ee
If the triple intersection point is an ordinary point in spacetime, then by zooming in on its vicinity, we can scale out the AdS radius. We then end up with three planes intersecting along lines which themselves move with the speed of light. Therefore the intersection point must move superluminally which is impossible.

\begin{figure}[h]
\begin{center}
\includegraphics[width=10cm]{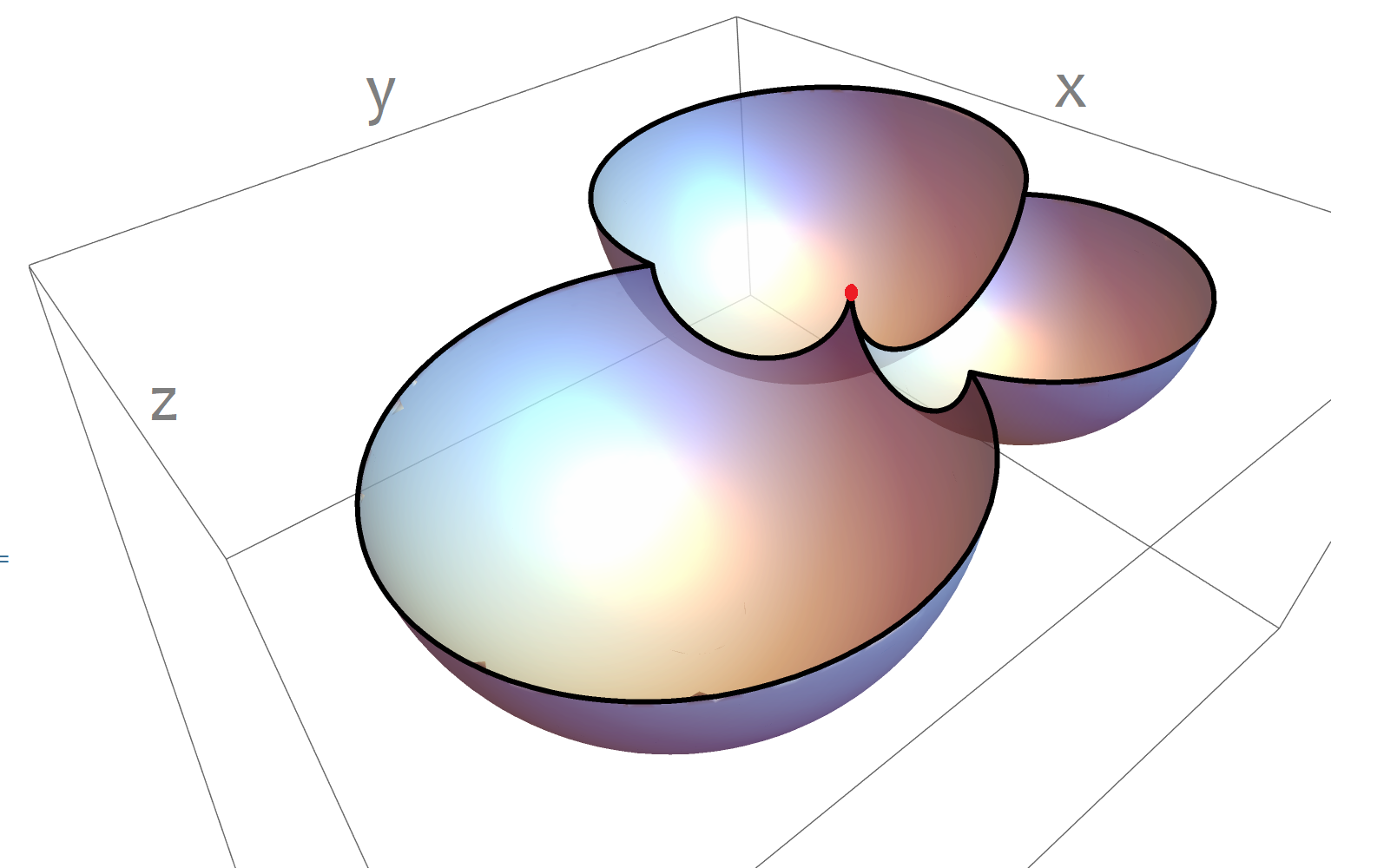}
\caption{\label{fig:trivalent}
Three compatible linear patches on the \poincare patch. The boundary of AdS is the plane on the top of the bounding box. The vertex (red dot) moves on the AdS boundary with the speed of light. The three semi-circles emanating from the vertex are the three shockwaves on the membrane.
}
\end{center}
\end{figure}

The only possibility is that the vertex moves on the boundary of AdS, see Figure \ref{fig:trivalent}.
Boundary points $X \in \RR^{2,3}$ are characterized by $X^2 = 0$ with the identification under rescaling $X \sim \lambda X$.
After an $SO(2,3)$ transformation the normal vectors can be written as
\bea
  \nonumber
  A &=& (0, \, 0, \, 0, \, 0, \, 1)^T \\
  \nonumber
  B &=& (0, \,  a, \,  -a, \,  0,  \, 1 )^T \\
  \nonumber
  C &=& (b, \,  c, \,  -c, \,  b,  \, 1 )^T
\eea
where $a,b,c$ are parameters.
The triple intersection locus is a one-dimensional subspace. Points on it satisfy
\be
  \label{eq:tripeqns}
  X\cdot A = 0 \qquad X\cdot B = 0 \qquad X\cdot C = 0 \qquad  X^2 = 0
\ee

Points on the boundary can be parametrized as
\be
  \nonumber
  X = \le(\half(-1+t^2-x^2-y^2), \, t, \, x, \, y, \, \half(1+t^2-x^2-y^2) \ri)^T
\ee
Eqn. (\ref{eq:tripeqns}) can be solved by setting $x(t) = -t$ and $y(t) = -1$. Hence, the vertex moves in the $-x$ direction with the speed of light.

By adding extra vertices and additional compatible membrane patches one could construct more complex membranes. The projection of the shockwaves on the $z=0$ boundary is then a trivalent graph (with linear edges).
In order to compute the global time-evolution, one would need to understand what happens when two trivalent vertices collide.

\clearpage

\begin{figure}[h]
\begin{center}
\includegraphics[width=8cm]{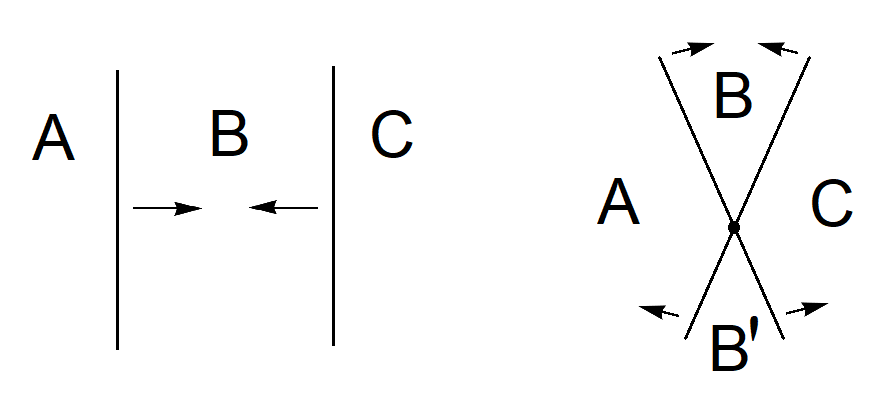}
\caption{\label{fig:boosting} {\it Left:} Schematic picture: a snapshot of the membrane which consists of three linear patches. The normal vectors are denoted by $A,B, C$. The two parallel shockwaves are moving towards each other.   {\it Right:} A boost of the system on the left in the vertical (parallel) direction produces crossing shockwaves. The vertex where the shockwaves cross moves superluminally.
}
\end{center}
\end{figure}

\section{Crossing shockwaves}

In the previous section we discussed that a hypothetical vertex connecting shockwaves between linear membrane patches necessarily moves faster than the speed of light if it is located in the bulk of AdS. In order to understand whether it is possible to construct a brane configuration with such a vertex, let us first look at the case of segmented strings. Let us assume that the string consists of three segments, with two kinks moving towards each other. Let us denote the normal vectors by $A,B, C \in \RR^{2,2}$, where $B$ corresponds to the middle segment.  Compatibility requires $ A \cdot B = B\cdot C = 1$. This ensures that the kinks move on null geodesics.
When they collide, $B$ will jump according to the {\it reflection formula} \cite{Vegh:2015ska, Callebaut:2015fsa}
\be
  \label{eq:reflection}
   B' = - B + 4 { A + C \over ( A +  C)^2 }
\ee
It is easy to check that $  A \cdot  D =  D \cdot  C = 1$ after the collision.

Let us now add a transverse dimension to the string (and to the target space). The configuration is depicted in Figure~\ref{fig:boosting} on the left-hand side. (The figure is schematic because the shockwaves are drawn as straight lines even though lie in AdS.) After the shockwaves pass through each other, the middle patch will have a new normal vector, again computed by (\ref{eq:reflection}) where it is now understood that the vectors all lie in $\RR^{2,3}$.  If the membrane is boosted in the vertical direction which is parallel to the shockwaves, then the result will be the one on the right-hand side. Now on the same time-slice all four patches can be seen, including the one corresponding to $B'$.

\begin{figure}[h]
\begin{center}
\includegraphics[width=7cm]{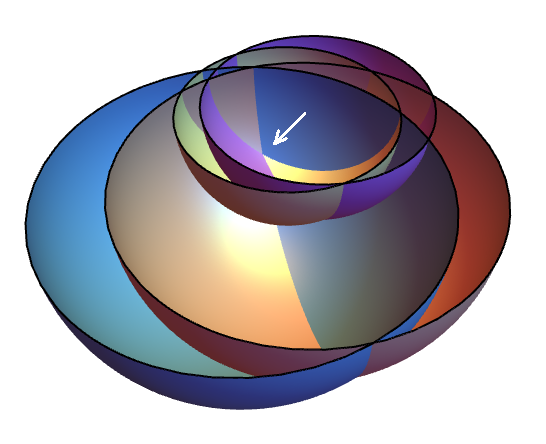}
\caption{\label{fig:4bubbles} Degree four vertex on a timeslice of the \poincare patch. It lies at the intersection of four hemispheres (the four linear membrane patches which are connected by the reflection formula as explained in the main text).
}
\end{center}
\end{figure}

This idea allows us to construct brane configurations with crossing shockwaves. Adjacent linear patches must be compatible and the four patches should be connected by the reflection formula (\ref{eq:reflection}).
An example on the \poincare patch of AdS is shown in Figure~\ref{fig:4bubbles} (with $\kappa=0$). The hemispheres of the four linear patches are shown in full. They touch the AdS boundary along their equators (black circles). White arrow indicates the location of the superluminal vertex where the shockwaves cross. The membrane solution consists of the four patches (blue, orange, purple and gray) around it.

More complex initial conditions for membranes can be constructed by adding further shockwaves to the setup. In order for the shockwaves to move with the speed of light, adjacent linear patches must be compatible and patches around four-valence vertices must be connected by the reflection formula.

\clearpage

\begin{figure}[h]
\begin{center}
\includegraphics[width=16cm]{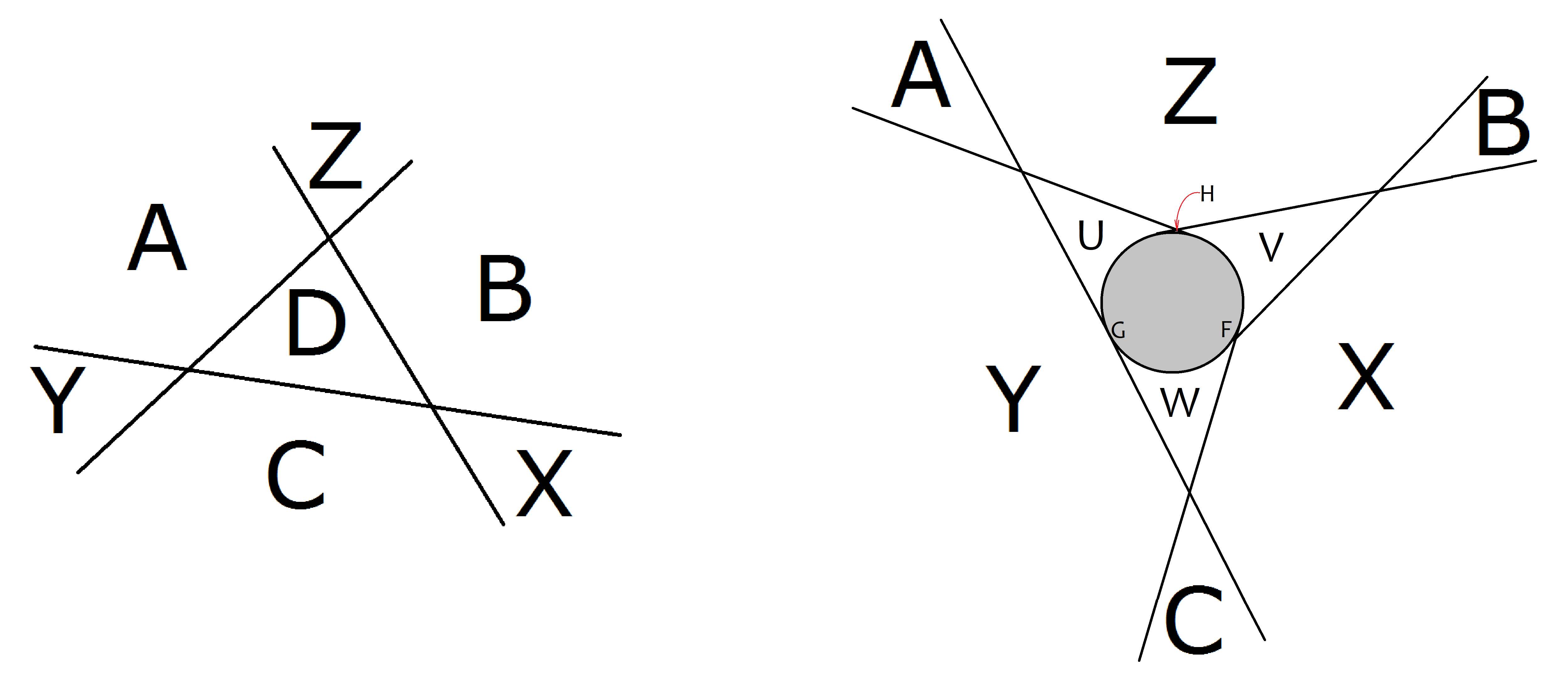}
\caption{\label{fig:three_shocks} Schematic picture of the collision of three shockwaves on the membrane. {\it Left:} Before collision the membrane consists of seven linear patches. {\it Right:} After collision, a non-linear patch forms in the middle (indicated by a gray disk). The boundary of this patch (black circle) lies on the intersection of the membrane and the future lightcone of the collision event. Normal vectors outside the lightcone are dictated by the reflection formula.
}
\end{center}
\end{figure}

\section{Collision of three shockwaves}

In the previous section we discussed how piecewise linear membrane configurations can be built in AdS. A natural question arises: what happens when three shockwaves collide and a patch vanishes?

The situation before collision is shown in Figure~\ref{fig:three_shocks} on the left-hand side. The three shockwaves on the membrane are moving towards each other and the size of $D$ decreases (with a slight abuse of notation the same letter will be used for patches and their normal vectors).
Preliminary numerical simulations show that after collision the membrane will not be linear any longer. This is seen on the right-hand side of the figure. There are several new linear patches on the brane: $U, V, W$ and three smaller ones $F, G, H$. These are all computed by the reflection formula from three adjacent patches around a vertex. For instance, $U$ can be computed from $A, Z$ and $Y$. In the middle of the membrane a non-linear patch is formed. This is indicated by a gray disk in the figure.  The boundary of this patch (indicated by a black circle) lies on the future lightcone of the collision event. Inside the gray region the membrane `knows' about the collision and interactions deform its shape.

\clearpage

\section{Discussion}

In this paper we discussed membrane embeddings in AdS$_4$  which satisfy the equations of motion of a Nambu-Goto membrane (coupled to a background three-form). Various membrane solutions can be obtained by gluing linear patches. Since the brane is a codimension one object in this target space, patches can be characterized by their constant normal vectors. The curve where adjacent patches intersect is a shockwave on the membrane. It has to propagate with the speed of light, otherwise the brane will bend and lose its piecewise linear nature. This imposes a compatibility condition on normal vectors of adjacent patches.
We have also discussed cases with crossing shockwaves where the intersection point was either on the boundary of AdS (trivalent vertices) or in the bulk (valence-four vertices).
Although an underlying { three-dimensional} discrete integrable theory is not present here, some of these solutions can be described by a { two-dimensional} discrete integrable theory.
Although we have not discussed how to glue quadric patches, it also seems possible. For  a similar construction in higher dimension, see \cite{Fiol:2014vqa}.

\vspace{0.2in}   \centerline{\bf{Acknowledgments}} \vspace{0.2in}
The author is supported by the STFC Ernest Rutherford grant ST/P004334/1.

\bibliographystyle{JHEP}
\bibliography{broken}

\end{document}